\documentclass[twocolumn,showpacs,prc]{revtex4}       

\newcommand{ \be }{\begin{equation}}    
\newcommand{ \ee }{\end{equation}}    
\newcommand{ \bea }{\begin{eqnarray}}    
\newcommand{ \eea }{\end{eqnarray}}

\usepackage{graphicx}
\usepackage{dcolumn}
\usepackage{bm}
\usepackage{verbatim}
\usepackage{amsmath}

\begin{document}       
\voffset=0.5 in    
       
       
\title{    
Effects of the Detection Efficiency on Multiplicity Distributions
}    
       
\author{A.H. Tang$^1$ and G.Wang$^2$}
\affiliation{$^1$Brookhaven National Laboratory, Upton, New York 11973 \\
$^2$University of California, Los Angeles, California 90095 \\}
       
    
\begin{abstract}       
In this paper we investigate how a finite detection efficiency affects three popular multiplicity distributions, 
namely the Poisson, the Binomial and the Negative Binomial distributions. 
We found that a multiplicity-independent detection efficiency does not change the characteristic of a distribution,
while a multiplicity-dependent detection efficiency does. 
We layout a procedure to study the deviation of moments and their derivative quantities from the baseline distribution due to a multiplicity-dependent detection efficiency. 
\end{abstract}       
       
    
\maketitle       
       
\section{\label{sec:level1}INTRODUCTION}       
       
One of the main purposes of relativistic heavy ion collision experiments is to explore the QCD phase boundary~\cite{whitePapers}, 
in particular to look for signatures of a first order phase transition~\cite{Ejiri,Bowman} and a critical end point~\cite{Stephanov,Fodor}. 
Moments of the distributions of conserved quantities, such as net-baryon number, net-charge and net-strangeness, 
have been argued to 
be sensitive to the phase transition and the critical end point, and are drawing increased attention from both experimentalists
\cite{STARNetProtonPRL,STARNetCharge,PhenixNetCharge} and theorists~\cite{Stephanov09,Asakawa09,Athanasiou10,Stephanov11}. 
In the study of higher order moments and their derivative quantities, an abnormal deviation from the baseline distribution is 
usually interpreted as an interesting physics signal. In practice, such a deviation is complicated by experimental effects, such as 
a finite detection efficiency. In this paper, we address how a finite efficiency would change three widely used multiplicity 
distributions, namely, the Poisson, the Binomial and the Negative Binomial distributions. We will discuss the case of a multiplicity-independent 
efficiency, followed by the case of a multiplicity-dependent efficiency, where we layout a procedure to investigate how the efficiency affects the 
three multiplicity distributions. The procedure also applies to the difference distribution of two multiplicity distributions. 
     
\section{\label{sec:multDistrib} Multiplicity Distributions with a Multiplicity Independent Efficiency}       
       
\subsection{Poisson Distribution}        

The probability mass function for the Poisson distribution is given by
\be   
f(k;\lambda)= \frac{\lambda^k e^{-\lambda}}{k!},
\label{eq:Poisson}
\ee   
where $k$ is a non-negative integer (same for the other two distributions discussed below), 
and $\lambda$ is both the mean and the variance of the distribution.  
The probability-generating function for the Poisson distribution is given by

\be   
G(z)= e^{-\lambda(1-z)},
\label{eq:PoissonGenFcn}
\ee 
where $z$ is a complex number with $|z|\le1$.

We treat observing and not-observing a particle as ``decay'' modes of a particle, and apply the cluster decay 
theorem~\cite{Pumplin} by replacing $z$ with the generating function 

\be   
g(y)= (1-\epsilon)+\epsilon y,
\label{eq:singleGFcn}
\ee 
where $\epsilon$ is the probability of seeing a particle, in practice less than unity due to the finite acceptance 
and detection efficiency. Without losing generality, below we refer to $\epsilon$ as the detection efficiency inclusive of both sources.

Then Eq.~(\ref{eq:PoissonGenFcn}) becomes

\bea  
G(z) = G(g(y)) &=& e^{-\lambda\big(1-[(1-\epsilon)+\epsilon y]\big)} \nonumber \\ 
        &=& e^{-\lambda\epsilon(1-y)}.
\label{eq:PoissonGenFcnModified}       
\eea       

One immediately identifies that the new generating function, for an experimental observable with a finite detection 
efficiency, still maintains the form of a Poisson distribution, with the mean of the distribution reduced to $\lambda \epsilon$.

Note that Eq.~(\ref{eq:singleGFcn}) is simply the generating function for a Binomial process with $n=1$ (see below). 
With Eq.(\ref{eq:singleGFcn}) convoluted into Eq.(\ref{eq:PoissonGenFcn}), the fluctuation of event-by-event efficiency has been taken into account, similar to the procedure proposed in~\cite{Adam}.

\subsection{Binomial Distribution}        

The probability mass function for the Binomial distribution is given by       
\be   
f(k;n,p)= {n \choose k} p^k (1-p)^{n-k},
\label{eq:Binom}
\ee   
where $p \in [0,1]$, and the non-negative integer $n \ge k$. 
The mean and the variance of the distribution are given by $np$ and $np(1-p)$, respectively. 
The corresponding probability-generating function is given by 
\be   
G(z)= (1-p+pz)^n.
\label{eq:BinomGenFcn}
\ee 

Similarly with a finite detection efficiency, 

\bea  
G(z) = G(g(y)) &=& [1-p+p(1-\epsilon+\epsilon y)]^n \nonumber \\
                       &=& [1-p\epsilon + p\epsilon y]^n.
\label{eq:BinomGenFcnModified}       
\eea    

We have recovered the probability-generating function for the Binomial distribution with the replacement 
of $p \rightarrow p' (=p\epsilon)$. The mean of the new distribution is given by $\mu'=\mu\epsilon$. 
The calculation of other quantities under the influence of a finite detection efficiency is thus straightforward.
For example,
\bea
\kappa\sigma^2 = \frac{C_4}{C_2}=1-6p+6p^2,
\eea
where $\kappa$ is the kurtosis and $C_i$ is the $i$th order cumulant.
When taking the detection efficiency into account, one simply replaces every $p$ with $p\epsilon$,
\bea
\kappa\sigma^2 = \frac{C_4}{C_2}=1-6p\epsilon+6p^2\epsilon^2.
\eea
Such knowledge is useful for quantifying the deviation of the observable of interest from the original distribution 
due to the finite detection efficiency.

\subsection{Negative Binomial Distribution}        

The probability mass function for the Negative Binomial distribution is given by
\be   
f(k;r,p)= {k+r-1 \choose k}(1-p)^k p^r,
\label{eq:NegBinom}
\ee   
where $p \in [0,1]$, and the real number $r>0$. 
It has identities of $p=\frac{\mu}{\sigma^2}$ and $r=\frac{\mu p}{1-p}$, where $\mu$ and $\sigma^2$ are the mean and the variance, respectively. Its probability-generating function has the form of
\bea   
G(z) &=& \left( \vphantom{} \frac{\frac{r}{\mu}}{1+\frac{r}{\mu}-z} \right)^r \nonumber \\
        &=& \left( \vphantom{} \frac{p}{1-(1-p)z} \right)^r,
\label{eq:NegBinomGenFcn}
\eea   
where $p=\frac{\mu}{\sigma^2}=\frac{r}{\mu+r}$.

Likewise, in the case of a finite detection efficiency, we have
\bea  
G(z) = G(g(y)) &=&\left( \vphantom{} \frac{p}{1-(1-p)(1-\epsilon+\epsilon y)} \right)^r \nonumber \\
		      &=& \left( \vphantom{} \frac{p'}{1-(1-p')y} \right)^r,
\label{eq:BinomGenFcnModified}       
\eea    
where $p'=\frac{p}{\epsilon+p-p\epsilon}$, and $r$ is unchanged. 
The form of the probability-generating function for the Negative Binomial distribution is recovered, 
with $p \rightarrow p'$ and $\mu \rightarrow \mu'(=\mu\epsilon)$.
Again, other quantities with a finite detection efficiency can be evaluated with the two simple replacements. 
For example, replacing $p$ with  $\frac{p}{\epsilon+p-p\epsilon}$ everywhere in
\bea
\kappa\sigma^2 = \frac{C_4}{C_2}=\frac{6-6p+p^2}{p^2}
\eea
gives the $\kappa\sigma^2$ for the case with a finite detection efficiency.

\section{\label{sec:multDistrib} Multiplicity Distributions with a Multiplicity Dependent Efficiency}       
Usually the detection efficiency decreases with increased multiplicity, as the reconstruction of a particle becomes more 
difficult in a crowded environment. In this case, the detection efficiency is expressed
as a function of $k$, $\epsilon(k)$. Now for all the three distributions, 
the probability-generating function can no longer be written 
in a concise form. Instead, we take the general definition
\bea  
G(y) &=& \sum\limits_{k=0}^{\infty}f(k) z^k \nonumber \\ 
        &=& \sum\limits_{k=0}^{\infty}f(k) [1-\epsilon(k)+\epsilon(k)y]^k.
\label{eq:GenFcnModified2}       
\eea

Generally one cannot recover the generating function of the same type. That means, 
a multiplicity-dependent efficiency will distort the original distribution,
unlike the case of a multiplicity-independent efficiency, where the detector effect will change the mean and width of the distribution,
but keep the characteristic shape (as the same type). 
Nevertheless, with $\epsilon(k)$ as input, one can still calculate the mean ($\mu'$) and the variance ($\sigma'^2$):
\be 
\mu'= \langle M \rangle = F_1,
\ee
\bea
\sigma'^2 &=& \langle M^2 \rangle - \langle M \rangle^2 \nonumber \\
                &=& \langle M(M-1) \rangle + \langle M \rangle - \langle M \rangle ^2 \nonumber \\
                &=& F_2 + F_1 - F_1^2,
\eea
where $F_i$ is the factorial moment $\langle M(M-1)\cdots(M-i+1)\rangle$, 
given by $F_i\equiv \frac{\partial^i G(y)}{\partial y^i} \bigg|_{y=1} $. 

For the Poisson distribution
\be 
F_i= e^{-\lambda}\sum\limits_{k=i}^{\infty}\frac{\lambda^k}{(k-i)!}\epsilon(k)^i ,
\label{eq:Fi_diffGPoisson}
\ee
for the Binomial distribution
\be
F_i= \sum\limits_{k=i}^{\infty} \frac{n!}{(k-i)!(n-k)!}p^k(1-p)^{n-k}\epsilon(k)^i ,
\label{eq:Fi_diffGBinomial}
\ee
and for the Negative Binomial distribution
\be
F_i= \sum\limits_{k=i}^{\infty} \frac{(k+r-1)!}{(k-i)!(r-1)!}(1-p)^kp^r\epsilon(k)^i.
\label{eq:Fi_diffGNegBinomial}
\ee
With Eq. (\ref{eq:Fi_diffGPoisson}), (\ref{eq:Fi_diffGBinomial}) and (\ref{eq:Fi_diffGNegBinomial}),  
$F_i$ can be numerically calculated with known $\epsilon(k)$, and the calculation is no more complicated than that for 
the corresponding distributions with the perfect detection. 
Note that in practice one only needs to perform the summation over $k$ to a value that is large enough, say, a few $\sigma$ above 
the mean value, so that $F_i$ has little change with further increase of $k$~\cite{footNote1}. 

The third and the fourth central moments are given by
\be
\big\langle (M - \langle M \rangle)^3 \big\rangle = F_1+2F_1^3+ 3F_2-3F_1(F_1+F_2)+F_3, 
\ee
and
\be
\begin{split}
\big\langle (M - \langle M \rangle)^4 \big\rangle = F_1-3F_1^4+ 7F_2+ 6F_1^2(F_1+F_2) + \\ 6F_3-4F_1(F_1+3F_2+F_3)+F_4.
\end{split}
\ee

With the mean, the variance, the third and the fourth central moments, the first few cumulants can be calculated as usual
\bea
C_1 &=& \langle (\delta M) \rangle = 0 \nonumber \\
C_2 &=& \langle (\delta M)^2 \rangle \nonumber \\
C_3 &=& \langle (\delta M)^3 \rangle \nonumber \\
C_4 &=& \langle (\delta M)^4 \rangle - 3 \langle (\delta M)^2 \rangle^2,
\eea
where $\delta M = M - \langle M \rangle$. One can further calculate skewness and kurtosis based on cumulants, which is straightforward and thus is not repeated here. 

Note that although we addressed three specific multiplicity distributions, the procedure discussed in this section can be extended to other multiplicity distributions, as long as the factorial moments can be conveniently calculated.

\section{\label{sec:diffMultDistrib} Difference Distribution of Two Multiplicity Distributions}       
       
The difference between two independent variables is useful for studying the fluctuation of conserved quantities, 
e.g. the net charge and the net baryon number. The difference between two variables, each following the Poission distribution,
is called the Skellam distribution, and its probability-generating function is given by:
\be
G(z;\mu_1,\mu_2)=e^{-(\mu_1+\mu_2)+\mu_1 z + \mu_2 / z}.
\label{eq:SkellamGFcn}
\ee
It follows from one of the properties of the probability-generating function:
for the difference of two independent random variables $S=X_1-X_2$, 
the generating function is given by $G_S(z)=G_{X_1}(z)G_{X_2}(z^{-1})$.
The generating function for the difference between two Binomial variables is
\be   
G(z;n_1,p_1,n_2,p_2)= (1-p_1+p_1 z)^{n_1}(1-p_2+p_2 / z)^{n_2},
\label{eq:DiffBinomGFcn}
\ee 
and the generating function for the difference between two Negative Binomial variables is
\be   
G(z;r_1,p_1,r_2,p_2) = \left( \vphantom{} \frac{p_1}{1-(1-p_1)z} \right)^{r_1} \left( \vphantom{} \frac{p_2}{1-(1-p_2)/z} \right)^{r_2}. 
\label{eq:NegBinomGenFcn}
\ee 
When we take into account the finite detection efficiency, none of the three generating functions above can recover the form 
of the same type. Fortunately, they describe the \emph{difference between two quantities}, to both of which the argument on
the detection efficiency still applies. This facilitates the calculation of cumulants of the three difference-distributions
with the finite detection efficiency under consideration.
For example, for the net charge distribution, the additivity of cumulants directly gives $C_{\rm \Delta charge} = C_{+}-C_{-}$,
where $C_{+}$ and $C_{-}$ are cumulants for positively and negatively charged particles, respectively. 
The $C_{\rm \Delta charge}$ with a finite detection efficiency can be calculated this way as long as 
the distributions of separate charges are independent of each other. 
Here we assume that the two underlying distributions are completely independent of each other,
to solely investigate how a non-physics effect (finite detection efficiency) disturbs 
the baseline distribution, when studying cumulants of the difference of two variables.
This treatment is different from that in~\cite{Adam} where the derivation starts from the cumulants of the difference distribution,
with the correlation between the two variables already taken into account.

\section{\label{sec:conclusion} Conclusion}       

We have shown that for the Poisson, the Binomial and the Negative Binomial distributions, 
a multiplicity-independent efficiency will modify the mean and the width of the original distribution, 
but it does \emph{not} change the distribution type. With a known multiplicity-indenpendent efficiency, the original distribution
can be completely reconstructed from the measured one, and vice versa. 
However, a multiplicity-dependent efficiency will distort the original distribution. In this case it is difficult to recover the original distribution. Nevertheless, one can still study how a finite, multiplicity-dependent detection efficiency changes the original distribution. The procedure applies also to the difference distribution 
of two independent distributions.
With a known form of $\epsilon(k)$, the deviation of moments and their derivative quantities from the baseline distributions, 
can be estimated following the procedure presented in this paper.  Knowledge obtained in this work will help avoid the mis-interpretation of certain observables as signals of the possible phase 
transition and/or the critical end point.

\begin{acknowledgments}
The work for the case of a multiplicity-dependent detection efficiency was stimulated by a discussion with J. Dunlop, P. Sorensen and H. Wang. We thank A. Bzdak and V. Koch for a fruitful discussion. A.T. was supported by the US Department of Energy under Grants No. DE-AC02-98CH10886 and No. DE-FG02-89ER40531. G.W. was supported by the US Department of Energy under Grants No. DE-FG02-88ER40424.
\end{acknowledgments}


\begin{thebibliography}{99}
\bibitem{footNote1}
To avoid the problem of the overflowing with large numerical numbers, one may also calculate, for example,  $e^{-\lambda}\sum\limits_{k=i}^{\infty}\frac{\lambda^k}{(k-i)!}\epsilon(k)^i$ as $\mathrm{exp}\{\mathrm{ln}[\sum\limits_{k=i}^{\infty}\frac{\lambda^k}{(k-i)!}\epsilon(k)^i]-\lambda\}$.
\end{thebibliography}
\end{document}